\documentclass[a4paper,12pt]{article}
\pdfoutput=1
\usepackage{feynmp-auto,expdlist}
\usepackage{amsmath, amsfonts, amssymb}
\usepackage{graphicx}
\usepackage{enumerate}
\usepackage{hyperref}
\usepackage{latexsym}
\usepackage{dsfont}
\usepackage{hepnicenames}
\usepackage{enumerate}
\usepackage{soul}
\usepackage[normalem]{ulem}
\usepackage{comment}

\usepackage{units}

\usepackage{soul}

\usepackage{mathrsfs,graphicx,rotating,amsmath,amsfonts,mathtools,booktabs,amssymb,wasysym}
\usepackage{hyperref}\usepackage{slashed}
\usepackage[nosort]{cite}
\usepackage[table,xcdraw,dvipsnames]{xcolor}
\usepackage{bm}
\usepackage{graphicx}
\usepackage{multirow,multicol}
\hypersetup{colorlinks,bookmarksopen,bookmarksnumbered,
	linkcolor=blus,pdfstartview=FitH,urlcolor=rossos,citecolor=verde}
\allowdisplaybreaks

\renewcommand{\[}{\left[}

\newcommand{\Lag}{\mathscr{L}}

\newcommand{\mio}[1]{}

\newcommand{\bpm}{\begin{pmatrix}}
\newcommand{\epm}{\end{pmatrix}}

\usepackage{mathrsfs}

\newcommand{\fig}[1]{~\ref{fig:#1}}
\newcommand{\sfrac}[2]{#1/#2}

\allowdisplaybreaks
\usepackage{multicol}
\usepackage{color}
\definecolor{rosso}{cmyk}{0,1,1,0.4}
\definecolor{rossos}{cmyk}{0,1,1,0.55}
\definecolor{rossoc}{cmyk}{0,1,1,0.2}
\definecolor{blu}{cmyk}{1,1,0,0.3}
\definecolor{blus}{cmyk}{1,1,0,0.6}
\definecolor{bluc}{cmyk}{1,1,0,0.1}
\definecolor{verde}{cmyk}{0.92,0,0.59,0.25}
\definecolor{verdec}{cmyk}{0.92,0,0.59,0.15}
\definecolor{verdes}{cmyk}{0.92,0,0.59,0.4}

\newcommand{\bp}{\bar{M}_{\rm Pl}}
\newcommand{\eq}[1]{~{\rm (\ref{eq:#1})}}

\newcommand{\beq}{\begin{equation}}
\newcommand{\eeq}{\end{equation}}

\newcommand{\bea}{\begin{eqnarray}}
\newcommand{\eea}{\end{eqnarray}}
\newcommand{\be}{\begin{equation}}
\newcommand{\ee}{\end{equation}}
\font\tenrsfs=rsfs10 at 12pt
\font\sevenrsfs=rsfs7
\font\fiversfs=rsfs5
\newfam\rsfsfam
\textfont\rsfsfam=\tenrsfs
\scriptfont\rsfsfam=\sevenrsfs
\scriptscriptfont\rsfsfam=\fiversfs

\newsavebox\MBox

\renewenvironment{thebibliography}[1]
{\begin{multicols}{2}[\section*{\refname}]%
		\@mkboth{\MakeUppercase\refname}{\MakeUppercase\refname}%
		\list{\@biblabel{\@arabic\c@enumiv}}%
		{\settowidth\labelwidth{\@biblabel{#1}}%
			\leftmargin\labelwidth
			\advance\leftmargin\labelsep
			\@openbib@code
			\usecounter{enumiv}%
			\let\p@enumiv\@empty
			\renewcommand\theenumiv{\@arabic\c@enumiv}}%
		\sloppy
		\clubpenalty4000
		\@clubpenalty \clubpenalty
		\widowpenalty4000%
		\sfcode`\.\@m}
	{\renewcommand{\@noitemerr}
		{\@latex@warning{Empty `thebibliography' environment}}%
		\endlist\end{multicols}}

\newcommand{\SU}{\,{\rm SU}}
\newcommand{\SO}{\,{\rm SO}}

\makeatletter

\font\ital=cmu10

\newcommand{\hhref}[1]{\href{http://arxiv.org/abs/#1}{arXiv:#1}}
\usepackage{xstring}
\newcommand{\hhrefq}[1]{\IfSubStr{#1}{:}{\href{http://inspirehep.net/search?ln=en&ln=en&p=#1&of=hb&action_search=Search&sf=&so=d&rm=&rg=25&sc=0}{InSpire:#1}}{\hhref{#1}}}

\newcommand{\art}{\@ifnextchar[{\eart}{\oart}}
\newcommand{\article}{\@ifnextchar[{\earticle}{\oarticle}}
\newcommand{\oarticle}[7]{{\rm #1}, {\ital ``#6''}, {\rm #2 #3 (#5) #4}}

\newcommand{\doi}[1]{\href{http://dx.doi.org/#1}{[link]}}
\newcommand{\hhrefqq}[1]{\IfBeginWith{#1}{10.}{\href{https://doi.org/#1}{doi:#1}}{\hhrefq{#1}}}

\newcommand{\earticle}[7]{{\rm #2}, {\ital ``#7''}, {\rm #3 #4 (#6) #5}  [\hhrefqq{#1}]}

\renewenvironment{thebibliography}[1]
{\begin{multicols}{2}[\section*{\refname}]%
		\@mkboth{\MakeUppercase\refname}{\MakeUppercase\refname}%
		\list{\@biblabel{\@arabic\c@enumiv}}%
		{\settowidth\labelwidth{\@biblabel{#1}}%
			\leftmargin\labelwidth
			\advance\leftmargin\labelsep
			\@openbib@code
			\usecounter{enumiv}%
			\let\p@enumiv\@empty
			\renewcommand\theenumiv{\@arabic\c@enumiv}}%
		\sloppy
		\clubpenalty4000
		\@clubpenalty \clubpenalty
		\widowpenalty4000%
		\sfcode`\.\@m}
	{\renewcommand{\@noitemerr}
		{\@latex@warning{Empty `thebibliography' environment}}%
		\endlist\end{multicols}}

\newcounter{alphaequation}[equation]
\renewcommand{\thealphaequation}{\theequation\hbox to
	0.6em{\hfil\alph{alphaequation}\hfil}}
\definecolor{Gray}{gray}{0.95}

\usepackage{tocloft}
\begin{document}
\thispagestyle{empty}
\begin{center}  
{\Large\bf\color{rossos} Pole inflation from non-minimal coupling to gravity
} \\
\vspace{0.6cm}
{\bf Sotirios Karamitsos} and {\bf Alessandro Strumia}  \\[6mm]
{\it Dipartimento di Fisica, Universit\`a di Pisa, Pisa, Italia}\\[1mm]

\vspace{0.5cm}
{\large\bf Abstract}
\begin{quote}\large
Transforming canonical scalars to the Einstein frame
can give a multi-field generalization of pole inflation 
(namely, a scalar with a divergent kinetic term)
at vanishing field-dependent Planck mass.
However, to obtain an attractor, the scalar potential must obey certain non-generic conditions.
These are automatically satisfied in Quantum Field Theories with dimension-less couplings.
The resulting models of pole inflation have special
inflationary predictions determined by the full RG running of couplings.
Acceptable predictions for the tensor/scalar ratio arise for perturbative but moderately large couplings,
so we explore the possible QFT runnings:  to confinement, to an IR fixed point, and to a UV fixed point.

\end{quote}
\end{center}
\setcounter{page}{1}
\tableofcontents

\section{Introduction}

Predictive and acceptable models of inflation can be obtained under the assumption that
the Einstein-frame kinetic terms of some inflaton scalar contains a pole. A popular example of such models is $\alpha$--attractors \cite{1405.3646, 1502.07733, 1506.00936, 1505.03386,1504.00663,1703.00305}.
As reviewed in section~\ref{overview}, these models generate predictions that depend 
on the order and residue of the leading pole~\cite{1507.02277,1602.07867}.
The pole in the kinetic term is usually introduced by hand, motivated from SUSY-induced hyperbolic geometry \cite{1306.3214,1504.05557} or modified gravity~\cite{1612.01126,1802.06486}. However, an often overlooked way to motivate pole cosmology is through the inclusion of a vanishing conformal factor~\cite{1612.04730}.
Moreover, pole inflation can occur in the presence of multiple fields, but such scenarios have not been studied to a great extent in the literature.

\smallskip

In section~\ref{overview}, we provide an overview of pole inflation, and go on to consider more general tilts in the potential. This allows us to compute inflationary predictions
assuming that they do not depend on the detailed structure of the potential around the Standard Model vacuum 
(that is usually ignored in studies of pole inflation).
We briefly discuss how reheating can be achieved in such models.


\smallskip

In section~\ref{J2E}, we explore the possibility that Einstein-frame kinetic poles arise from 
Jordan-frame non-minimal  couplings of scalars to gravity in ordinary theories with no Jordan-frame kinetic poles.
Rewriting a generic theory with $N>1$ scalars in the Einstein frame, we find that
kinetic terms diverge along a $(N-1)$-dimensional surfaces in field space,
 \emph{singularity curves}.
 We find that a multi-field generalisation of pole inflation can arise, but only
 if the Jordan-frame scalar potential $V$ satisfies some non-generic condition.
Under this condition, a robust attractor profile remains even in the presence of additional fields, 
even with the added complication of multi-field effects.
In particular, even if the
field space is warped, this has little effect as its curvature does not generically diverge on the singularity curves.
We show that in many cases one obtains a 2nd order pole with a specific value of the residue,
which implies the same predictions of Starobinsky inflation.
We examine multi-field models with conic section singular surfaces. 

\medskip

In section~\ref{dimension-less}, we show that 
dimension-less theories provide a
quantum structure that produces pole inflation out of the origin of field space.
The  non-generic structure of the potential needed to obtain pole inflation is not postulated, but 
comes from combining general relativity with quantum mechanics.
Considering the simplest model with one scalar multiplet, we classify the inflationary predictions
depending on the quantum RG running of the couplings.
As we have full theories that can describe the structure of the potential around the SM minimum as well,
we relax the assumption that it is not relevant for pole inflation. This allows us to
obtain inflationary predictions different from usual pole inflation.
Small couplings universally lead to slow RG running and to (approximatively)
quadratic inflation, that is excluded because the predicted tensor/scalar ratio $r\approx 0.15$~\cite{1403.4226}
is now too large.
Moderate couplings add a higher order cubic term in the potential, 
which can reduce $r$ by an order unity factor down to an acceptable $r\approx 0.08$~\cite{1502.01334}.
Smaller values of $r$ need larger couplings and predictions tend to become model-dependent.
We explore the qualitatively different fast RG runnings that arise at larger couplings in QFT.
We find that a coupling that runs non-perturbative in the UV (Landau pole, section~\ref{Landau})
or in the IR (confinement, section~\ref{confinement}) can lead to small $r$.
On the other hand, couplings that run to an interacting fixed point in the IR (section~\ref{IRFP}) or 
in the UV (section~\ref{UVFP}) do not lead to particularly small $r$, at least in the examined
representative models.

\smallskip

Conclusions are given in section~\ref{concl}.

%

 \section{Overview of pole inflation}\label{overview}
In the Einstein frame, fields are parametrised such that scalars have minimal coupling to gravity.
Pole inflation is obtained when the Einstein-frame action of a scalar $\phi$
has the following form, with
non-minimal scalar kinetic term $K_{\rm E}(\phi)$:
\beq \label{eq:genericlagrangian}
S= \int d^4 x \sqrt{|\det g_{\rm E}|} 
\bigg[  - \frac{\bar M_{\rm Pl}^2}{2} R_{\rm E} + \frac{K_{\rm E}(\phi)}{2} {(D_\mu \phi)(D^\mu \phi)}-V_{\rm E}(\phi) + \cdots\bigg].\eeq
We here assume one scalar $\phi$  and (without loss of generality) a pole at $\phi=0$.
The Lagrangian close to the pole is usually assumed to be
\begin{align}
\label{eq:poleaction}
K_{\rm E}(\phi) \simeq  \frac{\alpha_p}{ |\phi|^p} ,\qquad
V_{\rm E}(\phi)\simeq V_{\rm E}(0) + V'_{\rm E}(0)\phi,
\end{align}
corresponding to the assumption that the pole in the kinetic term 
is not accompanied by a pole in the potential, that is approximated
by a first-order Taylor expansion.
We relax this assumption by assuming the slightly more general potential with 
a power $q$:
\beq\label{eq:Vpoleq}
 V_{\rm E}(\phi) \equiv V_{\rm E}(0) \left[1- \frac{(\beta \phi)^q}{q}\right].
\eeq
Rescaling the field $\phi\to \lambda \phi$ shows that physics only depends on the combination $\alpha_p \beta^{p-2}$.
The potential needs to have a negative gradient such that the field is pushed away from the pole,
towards the Standard Model (SM) minimum not described by the approximation of eq.\eq{poleaction} or\eq{Vpoleq}.

With one scalar only, the non-canonical  factor $K$ in the kinetic term for $\phi$ can be reabsorbed by defining a canonically normalised Einstein-frame scalar $\phi_{\rm E}(\phi)$ as $ {d\phi _{\rm E}}/{d\phi } = \sqrt{K}$.
The region near to the pole corresponds to an infinite range of the canonical $\phi_{\rm E}$ field
if $p\ge 2$,
\begin{align}
\label{eq:canoninfl}
\phi_{\rm E} =
\begin{cases}\displaystyle
 \frac{2\sqrt{\alpha_p} |\phi| ^{1-\frac{p}{2}}}{p-2} 	& \hbox{if }p \ne 2,\\ \displaystyle
 \sqrt{\alpha_2}  \ln  \left|\sfrac{\phi}{\bar{M}_{\rm Pl}}\right| 
& \hbox{if $p = 2$}.
\end{cases}
\end{align} 
This leads to a stretching of the potential and, in turn, to acceptable inflation if $q$ is low enough.
In the special dimension-less case $p=2$ the field range remains infinite also at large $|\phi|$.

For later reference we recall how the inflationary predictions can be computed without rewriting the action in terms of
a canonical scalar.
The slow-roll parameters\footnote{
Our slow-roll parameters are defined as $\epsilon= -\dot H/H^2$ and 
$\eta = \dot \epsilon/H\epsilon $.
Our $\eta = 2\tilde\eta-4\epsilon$ differs from the $\tilde\eta$ 
usually encountered in the literature, $\tilde\eta = \bar{M}_{\rm Pl}^2 V''/V$ for canonical kinetic terms. } 
are given by \cite{Burns:2016ric,Karamitsos:2017elm}
\beq
\label{eq:srp}
\epsilon \equiv \frac{1}{2} \frac{\bar{M}_{\rm Pl}^2}{ K_{\rm E}(\phi)} \frac{V'_{\rm E}(\phi)^2}{V_{\rm E}(\phi)^2} ,\qquad
\eta \equiv \frac{\epsilon'(\phi)}{\epsilon(\phi)} \frac{\bar{M}_{\rm Pl}^2}{K_{\rm E}(\phi)} \frac{V'_{\rm E}(\phi)}{V_{\rm E}(\phi)},
\eeq
and the number of $e$-foldings by
\begin{align}
\label{eq:efolds}
N (\phi) = \int_{\phi_{\rm end}}^{\phi }d\phi' \  \frac{K_{\rm E}(\phi')}{\bar{M}_{\rm Pl}^2} \frac{V_{\rm E}(\phi')}{V'_{\rm E}(\phi')}.
\end{align}
Inflation ends when $\epsilon(\phi_{\rm end}) \approx 1$.
The scalar tilt $n_s$, 
the amplitude of scalar perturbations $A_s$ and
the tensor-to-scalar ratio $r=A_t/A_s$ are given by the standard expressions
\begin{align}\label{eq:infpredth}
n_s = 1 -2 \epsilon+\eta,\qquad
A_s = \frac{1}{24\pi^2} \frac{V_{\rm E}}{\bar{M}_{\rm Pl}^4 \epsilon},\qquad
r = 16\epsilon
\end{align}
which are measured at about $N\approx 50-60$ $e$-folds before the end of inflation~\cite{2010.01139}
\beq \label{eq:infpredex}
n_s = 0.9649 \pm 0.0042 ,\qquad A_s \approx (2.1 \pm 0.06) \times 10^{-9},\qquad
r\lessapprox 0.044~(95\%\hbox{ C.L.}).
\eeq
Specialising to pole inflation, 
the number of $e$-folds close to the pole in eq.\eq{poleaction} becomes
\begin{align}
N \simeq  \frac{\alpha_p}{ \bar{M}_{\rm Pl}^2 \beta^q}
\begin{cases}\displaystyle
\frac{\phi^{2-p-q}-\phi_{\rm end}^{2-p-q}}{   p-2+q}  &  \hbox{if } p+q\ne2,
\\[2ex] \displaystyle
 \ln \frac{\phi_{\rm end} }{\phi} &  \hbox{if }p+q=2.
\end{cases}\qquad\hbox{where}\qquad
\phi_{\rm end} \approx \left(\frac{2 \alpha_p}{\bar M_{\rm Pl}^2\beta^{2q}}\right)^{\frac{1}{p-2+2q}}  .
\end{align}
Unlike $\phi_{\rm E}$, 
$N$ can be large even for $p<2$, provided that $q$ is small enough,
namely that the potential is flat enough.
The observables expressed in terms of $N$ are (see~\cite{1412.3797,1602.07867} for $q=1$)\footnote{In the special case $p+q=2$, where $c=\infty$, one instead has  
$ \label{eq:observables1}
n_s = 1 -\sfrac{q}{\tilde\alpha_p}$ and
$r \approx 16 e^{-qN/\tilde\alpha_p}$
where the pre-factor depends on the precise value of $\phi_{\rm end}$.}
\beq \label{eq:observables2}
n_s = 1 - \frac{c}{N},\qquad
r = 8  \left[ \frac{
\tilde\alpha_p^{q/(p-2+2q)} }{(p-2+q)  N } \right]^{c} ,\qquad
 c= \frac{p-2+2q}{p-2+q}\ge 1
\eeq
having defined the dimension-less combination $\tilde\alpha_p \equiv\alpha_p \beta^{p-2}/ \bar{M}_{\rm Pl}^2$.
Notice that $n_s-1$ is dominated by $\eta$, as it scales with a lower power of $N$ than $\epsilon$.
Data favour $p >1.5$ if $q=1$~\cite{1602.07867}. 
The case $p=2$ (also known as $\alpha$-attractor~\cite{1306.5220,1311.0472})
gives predictions for $n_s$ that do not depend on the parameter $q$ in the potential:
$n_s = 1-2/N$ and $r=8 \tilde\alpha_2/ (q N)^2$,
compatibly with current data
if $\tilde\alpha_2/q^2 \lessapprox20$.

\medskip

These predictions arise assuming that the structure of the potential around the SM minimum
negligibly affects inflation, that takes place in the region dominated by the pole.
In most realisations of pole inflation, approaching the singularity corresponds to climbing up a stretched plateau, and as a result, the field moves away from the singularity, effectively ending inflation. 
For a potential of the form in eq.~\eqref{eq:poleaction}, the field evolves away from
the pole. 
Graceful exit is achieved as long as the slow-roll parameter $\epsilon $ eventually surpasses unity. While the details are model-dependent, reheating can still occur generically: even if the field cannot oscillate in the shallow canonical potential, other mechanisms can contribute to the reheating of the universe, e.g.~instant preheating~\cite{hep-ph/9812289} or gravitational reheating~\cite{gravreheating}. After reheating, the canonical field will roll further down the potential, asymptotically coming to a stop due to Hubble drag, or by falling in  the SM vacuum.
This value of the field therefore finally sets the effective value of the Planck mass.

Finally, we discuss the opposite case where the scalar slowly rolls towards the pole, 
because the Einstein potential $V_{\rm E}$ is lower at the pole.
This case gives rise to eternal inflation, unless~$V_{\rm E}$  nearly vanishes at the pole.
Only in such a special case
eternal inflation is avoided, since eventually $\epsilon \approx 1$. 
Moreover, the pole remains inaccessible, since the canonical field is stretched to infinity. 
Reheating can then occur as described in above, followed by the stabilisation of the Planck mass as the canonical field gradually comes to a stop down the shallow $V_{\rm E}$.



\section{Multi-field pole inflation from graviton canonicalisation}\label{J2E}\label{sec:multifield} 
We consider $N$ scalars $\phi^i$ 
with no poles in their gauge-covariant kinetic terms 
and non-minimal  couplings to gravity
\beq \label{eq:Jordangeneric}
S = \int d^4x \sqrt{|\det g|} \,\bigg[ -\frac{1}{2} f(\phi) R + \frac{K_{ij}(\phi)}{2}(D_\mu \phi^i)(D^\mu \phi^j) - V(\phi)  +\cdots
\bigg] .\eeq
As is well known, this  ``Jordan frame'' action  can be brought to the Einstein frame where the graviton action is canonical
by the field redefinition $g^{\rm E}_{\mu\nu}= g_{\mu\nu}\times  f/\bar M_{\rm Pl}^2 $.

\smallskip

For $N=1$ scalar field  the action acquires the form of eq.\eq{genericlagrangian} with
\beq \label{eq:onefield}
K _{\rm E}= \bar M_{\rm Pl}^2\bigg(\frac{K_{11}}{f} +  \frac{3 f^{\prime 2}}{2 f^2}\bigg) ,\qquad
V_{\rm E}(\phi) =\frac{\bar M_{\rm Pl}^4 V}{f^2} ,\eeq
where $K_{11}$ can be assumed to be 1 since a one-dimensional field space can always be canonically normalized.
Assuming that $f$ is a generic function that crosses zero at some point
$f(\phi_*)=0$ (corresponding to strong gravitational interactions)
with non-vanishing derivative $f'(\phi_*)$, eq.\eq{onefield} results in a pole of order 2 at $\phi=\phi_*$.
However in general one has $V(\phi_*)\neq 0$, so $V_{\rm E}(\phi_*)=\infty$.
In order to have pole inflation the Einstein-frame potential must instead be finite at the pole:
one needs to assume an appropriate potential $V$ that vanishes quadratically fast at the pole.

\smallskip


For multiple scalar fields $\phi_i$, $i = 1,\ldots, N$,
the Einstein-frame action has the form 
\beq \label{eq:genericlagrangianmulti}
S= \int d^4 x \sqrt{|\det g_{\rm E}|} \bigg[  - \frac{\bar M_{\rm Pl}^2}{2} R_{\rm E} + \frac{K^{\rm E}_{ij}(\phi)}{2} {(D_\mu \phi^i)(D^\mu \phi^j)}-V_{\rm E}(\phi) + \cdots\bigg].\eeq
The presence of multiple scalar degrees of freedom leads to the concept of a \emph{field space}, which is equipped with a positive-definite \emph{field space metric}. 
This metric depends on the Jordan-frame metric $K_{ij}$ and
non-minimal scalar coupling to gravity $f(\phi)$, and is given by~\cite{1003.1159}
\begin{align}\label{eq:KEij}
K^{\rm E}_{ij} = \bar M_{\rm Pl}^2 \left( \frac{K_{ij}}{f} + \frac{3}{2}\frac{ 1}{f^2} 
\frac{\partial f}{\partial\phi^i}\frac{\partial f}{\partial\phi^j}\right),  
\end{align}
We will assume canonical kinetic terms in the Jordan frame, $K_{ij}=\delta_{ij}$. 
Even so,
the metric $K^{\rm E}_{ij}$ generically describes a warped field space, $d\phi^2 = K^{\rm E}_{ij}d\phi^i d\phi^j$.

A generic $f(\phi)$ can vanish along some surface, which we will refer to as a \emph{singularity curve}. 
This curve is the higher-dimensional analogue of a pole.
Generically the singularity curve is $N-1$ dimensional: we here assume this is the case,
and discuss in section~\ref{point} a special case of a singularity curve with lower dimension.

 Similar to the single-field case, the potential $V_{\rm E}$ along the singularity curve will be infinite unless the surface contains points where the Jordan-frame potential $V(\phi)$ vanishes quadratically fast, in which case the resulting $V_{\rm E}(\phi) = V/f^2$ is finite.
Even then, this is not enough to get pole inflation, because scalars that parametrise
the $f=0$ surface may lead to huge gradients of the $V_{\rm E}$ potential. 
Therefore, a non-generic condition remains necessary to get pole inflation:
defining $\phi_*$ as a local minimum of $V(\phi)$ restricted to the surface $f=0$,
one needs $V(\phi_*)=0$.

This condition is analogous to the vanishing of the cosmological constant, $V_{\rm E}(\phi_{\rm SM})=0$.
There may be a connection between the two issues.
In section~\ref{dimension-less}, we will present a possible rationale for  $V(\phi_*)=0$:
it is satisfied at the origin of field space in dimension-less theories.
Alternatively, one might conjecture that some fundamental theory needs to have $V(\phi_*)=0$, 
because it means that the regions of field space $f>0$ and $f<0$ are not physically separated by an infinite potential barrier,
and that masses cannot be much heavier than the field-dependent Planck scale $\sqrt{f}$.

If the above condition is satisfied, the fields that parametrise the $f=0$ surface become very heavy near to $\phi_*$,
and pole inflation proceeds along a one-dimensional `valley' in field space.

Before we examine the attractor nature of multi-field pole inflation, it is important to understand  how singularity curves affect the kinetic term of the theory and the evolution of trajectories within the field space. For a single-field theory, it is easy to see how poles cannot be crossed; indeed, the field would need to have infinite energy to do so. A more geometric way to see this is by observing that the field space distance $d\phi_{\rm E}^2 = K_{\rm E}(\phi)d\phi^2$ is infinite if the path of the field traverses a pole: even if the space is one-dimensional, it is not possible to normalise the field when its domain includes poles. As a result, a theory with poles is essentially a collection of multiple canonical theories that cannot communicate with each other \cite{1903.03707}.

In the case of multiple fields, the field space metric has $N(N+1)/2$ independent entries, but it does not suffice for one of these entries to go to infinity at a point to cause the singularity curve to act as a barrier for the field. Indeed, the singularity may be an artefact of the way we have parametrised the fields.
For example, consider the field space line element 
\begin{align}
\label{flse}
d\phi^2=   K_{11}(\phi_2)d\phi_1^2 +  \,  {d\phi_2^2}
\end{align}
It is indeed possible to cross the singularity curve $K_{11}(\phi_2) = \infty$ if throughout the entirety of the trajectory, $d\phi_1 = 0$ is satisfied. Therefore, this is only an apparent singularity. However, compare this with the line element of a different field-space metric:
\begin{align}
\label{flse2}
d\phi^2=   K_{11}(\phi_1)d\phi_1^2 +  \,  {d\phi_2^2}
\end{align}
In this case, it is not possible to cross the singularity $K_{11}(\phi_1) = \infty$ while maintaining $d\phi _1= 0$, which indicates that this curve can never be crossed.  
It is worth noting that if the source of the pole is from a vanishing non-minimal coupling, the resulting singularity  curve cannot be crossed. 
This can be seen by examining the line element: 
one can, at least locally, rewrite the field-space metric switching to 
one coordinate $\phi_1$ `perpendicular' to the $f=0$ surface
(for example $\phi_1 =f$ itself) plus  $n-1$ coordinates along which $f$ is constant.
Crossing $f=0$ needs varying $\phi_1$, but its metric element has a pole at $f=0$.
Whenever $f$ is approximatively linear around $f=0$, 
a dominant 2nd-order pole with residue $\tilde\alpha_2=3/2$
arises from  the second term in eq.\eq{KEij} 
\beq  K_{11}^{\rm E}(\phi_1)= \frac{3\bp^2}{2\phi_1^2}.\eeq

\begin{figure}[t]
$$
\includegraphics[width=0.42\textwidth]{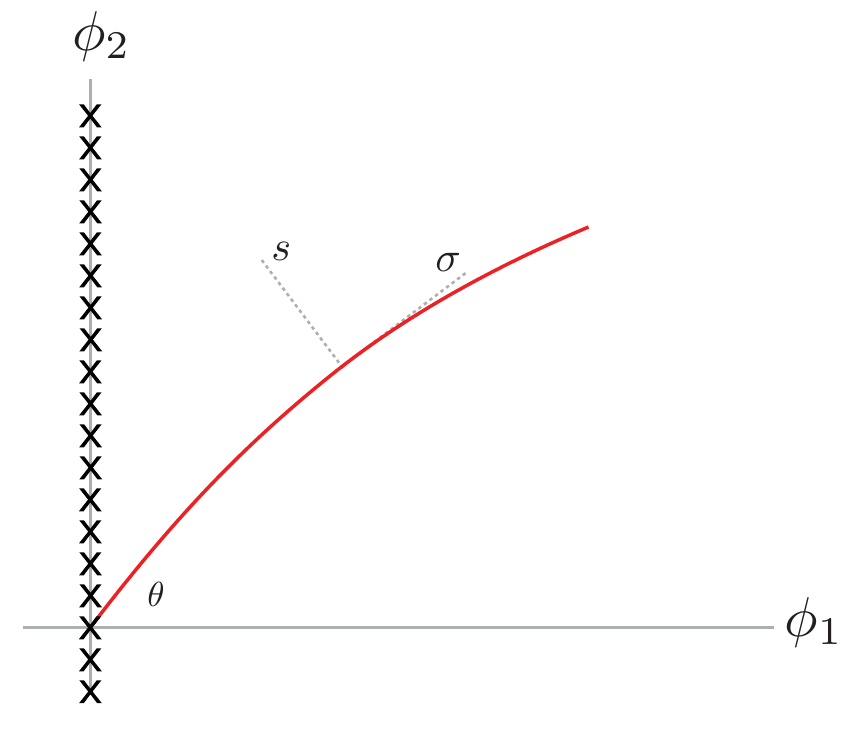} \qquad
\includegraphics[width=0.42\textwidth]{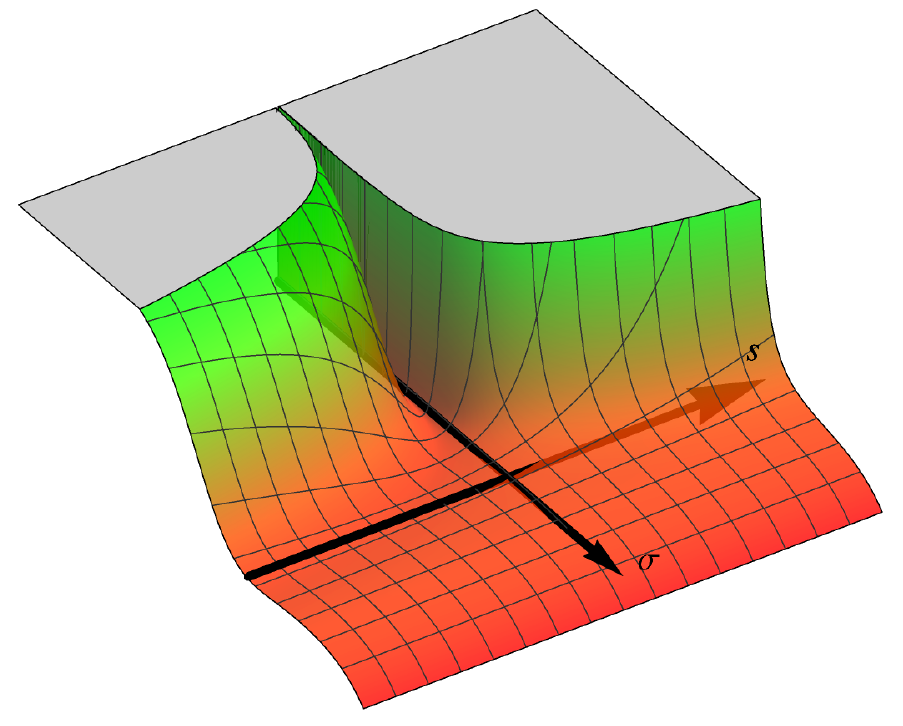} 
$$
\caption{\label{fig:trajectoryplot}\em  {\bfseries Left}: trajectory (red line) in two-field inflation in the neighbourhood near a singularity curve (dark crosses). 
 {\bfseries Right}: typical pole-induced inflationary potential: a narrow flat valley in canonical coordinates. }
\end{figure}

Let us next discuss the implications for inflation.
Since the components of $K_{ij}^{\rm E}$ diverge at $f=0$, one might worry that
the field-space curvature too can diverge  at $f=0$.
General arguments\footnote{The warped $N$-dimensional
field space can be described as a surface in a $N+1$-dimensional field space
with constant curvature, which leads to a geometric interpretation of the phenomenology of the theory \cite{1606.08848}. Indeed,
adding a $R^2/6f_0^2$ term to the Jordan-frame action, adds one extra scalar $z$ contained in the graviton,
and describing the overall scale factor in space-time.
This scalar becomes explicit going to the Einstein frame.
For $f\ge 0$, the $N+1$-dimensional field space is conformally flat with constant curvature
and a pole of order 2 at $z=0$~\cite{1502.01334}:
\beq \label{eq:LKin}
 \Lag_{\rm kin}^{\rm E} =   \frac{6\bp^2}{z^2}
 \frac{(D_\mu \varphi_i)^2 + (\partial_\mu z)^2}{2},\qquad
V_{\rm E}(z,\varphi)= \frac{36\bp^4}{z^4}
\bigg[{V(\varphi)}+   \frac{3f_0^2}{8} \bigg( f - \frac16 z^2 \bigg)^2\bigg] .\eeq
For large $f_0^2$ the extra field $z$ is heavy and can be integrated out as
$z^2=6f$, recovering the formulation
with $N$ scalars.
The $f=0$ curve separates the theory at $f>0$ from the theory with ghosts at $f<0$.} 
or explicit computations show that this is generically not the case.
Let us consider, for example,
a Jordan-frame action of the form in eq.\eq{Jordangeneric} with two scalars $\phi_1, \phi_2$
and coupling to gravity $f =M^2 + \xi_1 \phi_1^2 + \xi_2\phi_2^2$.
The  field space metric is 
\begin{align}\label{eq:fieldspacemetric}
K^{\rm E}_{ij} =  
\frac{\bp^2}{f^2}
\begin{pmatrix}
M^2 +\xi_1 (1+6\xi_1) \phi_1^2 + \xi_2 \phi_2^2 &  6 \bar{M}_{\rm Pl}^2 \xi_1\xi_2 \phi_1\phi_2\\
   6  \bar{M}_{\rm Pl}^2 \xi_1\xi_2 \phi_1\phi_2  &
M^2 +\xi_1 \phi_1^2 + \xi_2  (1+6\xi_2)\phi_2^2\\
\end{pmatrix}.
\end{align} 
The curvature in field space
diverges at $M^2 + \xi_1 (1+6\xi_1) \phi_1^2 + \xi_2(1+6\xi_2)\phi_2^2=0$,
away from $f=0$.
Then,
multi-field pole inflation does not significantly alter the nature of single-field pole inflation. 
To illustrate this, consider a simple theory with field space metric given in eq.~\eqref{flse2}. 
As discussed above, every theory with a $n-1$-dimensional
singularity curve can be cast in this form locally by a judicious choice of coordinate system. 
In view of our assumptions on the potential, and since field-space curvature is finite,
the inflationary trajectory has a low turn rate\footnote{Defined as the norm of the turn vector
 $
 \omega^i = \sfrac{d^2 \phi^i}{dNd\sigma}
 $
 where $\phi^i = \phi^i(\sigma)$ is the trajectory of the field,
 and $\sigma$ is the length in field space. 
Specific forms of the inflationary potential (for example, for $p=2$, a potential that depends on $\ln\phi_1$)
can lead to a non-negligible turn rate.}
in field space near the pole, and therefore approaches the singularity curve with 
some angle~$\theta$ with respect to the $\phi_1$-axis, as shown in fig.~\ref{fig:trajectoryplot}.
We change the coordinate system $(\phi_1,\phi_2,\ldots)$ to $(\sigma,s_1,\ldots)$,
where $\sigma$ is parallel to the inflationary trajectory and acts as the effective inflaton,
and $s_i$ are orthogonal.
A dominant pole given by $K^{\rm E}_{\rm 11}(\phi_1) = \alpha_p/ |\phi_1|^{p}$,
yields the following line element along the trajectory of the inflaton:
\begin{align}\label{twofieldelement}
d\phi^2|_{\rm pole} = \frac{ \alpha_p  }{|\sigma|^p} d\sigma^2 \, \cos^{2-p}\theta 
\end{align}
having neglected sub-leading poles.
The turn of the trajectory does not have an effect on the order of the pole, so the scalar tilt is unaffected. 
The $\cos^{2-p}\theta$ term affects the residue, and as a result can cause a \emph{suppression} on the scalar-to tensor ratio, 
more pronounced if the trajectory is oblique to the singularity curve.\footnote{Despite the notational similarity, this angle bears no relation to the so-called ``mixing angle'' (which depends on the transfer of isocurvature perturbations) \cite{astro-ph/0205253}.} 
The observables in eq.~\eqref{eq:observables2} are modified simply by making the substitution $\alpha_p \to \alpha_p \cos^{2-p} \theta$ assuming that the trajectory does not turn sharply in the direction of the singularity curve. 
A locally linear $f$ gives $p=2$, so that $\tilde\alpha_2 =3/2$ is unaffected by the angle $\theta$,
leading to the same predictions as Starobinsky inflation
\beq \label{eq:nsrpred}
n_s = 1- 2/N,\qquad r=12/(qN)^2\eeq
if the potential $V$ is approximatively linear, $q=1$.

Before we examine a few choice models, it is interesting to contend with the possible physical origins of singularities in the kinetic term that can lead to the behaviour described above. Kinetic poles are well established in string theory: they can be traced to K\"ahler potentials which arise generically due to string compactification \cite{kahler1, kahler2}. In such scenarios, additional moduli fields are not necessarily introduced. However, the fact that multiple fields contribute to inflation in our scenario (as opposed to acting like spectators) helps us draw parallels to N-flation \cite{nflation}. In fact, we find ourselves in a situation similar to \emph{pole} N-flation \cite{polenflation}, where the Einstein frame kinetic term achieves a second-order ellipsoidal pole (singular locus, in our language) for an arbitrary number of fields. In both scenarios, as the fields approach the singularity, one can identify a single variable that explicitly features a pole of second order, leading to pole inflation as usual. While our treatment so far has been restricted to two fields, it is not difficult to see that near the singularity, one degree of freedom would always feature a pole, even if more fields were added. In this sense, our approach incorporates pole N-flation, albeit without reference to the physical origin of the poles themselves.

We now turn our attention to the global structure of a few particular two-field models that demonstrate the attractor nature of inflation.

\subsection{Two-field pole inflation: hyperbolic singularity}
We consider a model with two fields $\phi_1 $ and $\phi_2$
and $f=\bp^2 +\xi_1 \phi_1^2 + \xi_2\phi_2^2$
assuming $\xi_1 = \xi_+>0$ and $\xi_2 =- \xi_-<0$ so that there is
a hyperbolic locus locus in which the kinetic term blows up. This singularity is generic: we can observe this by 
The field-space metric is obtained setting $M=\bp$ in eq.\eq{fieldspacemetric}. 
It is convenient to switch to elliptic coordinates ($\rho,\theta$):
\begin{align}
\label{ell1}
\phi_1 =  \bp \sqrt{\frac{2}{\xi_+}} \sinh \rho \sin(\pi/4 - \theta)   ,
\qquad
\phi_2= \bp \sqrt{\frac{2}{\xi_-}} \cosh\rho \cos( \pi/4 -  \theta) 
\end{align}
so that $f=\bp^2 \cosh(2\rho)\sin(2\theta)$,
the singularity locus is  $ \theta=0$, and
lines of constant $\theta$ are hyperbol\ae.
Only the $K^{\rm E}_{\theta\theta}$ component contains a 2nd-order pole,
and this is the only term relevant  for computing the observables:
\begin{align}
K^{\rm E}_{\theta\theta}|_{\rm pole}=\frac{3\bp^2}{2\theta^2}.
\end{align}
It is important to note that the linear vanishing of the non-minimal coupling $f$ in these coordinates is not a coordinate artifact, as we can verify that at the singularity locus, the first derivative of the kinetic term with respect to the original field $\phi_1$ still goes to infinity:
\begin{align}
K_{,\phi_1}|_{\rm pole} =  \frac{1}{M_P^2 + \xi_+ \phi_1^2 + \xi_- \phi_2^2}
\begin{pmatrix} 
 -24 \xi_+^3 \phi_1^3 & 24 \xi_+^2 \sqrt{\xi_-} \phi_1^2 \sqrt{M_P^2+\xi_+ \phi_1^2} \\
 24 \xi_+^2 \sqrt{\xi_-} \phi_1^2 \sqrt{M_P^2+\xi_+ \phi_1^2} & -24 \xi_+ \xi_- \phi_1 \left(M_P^2+\xi_+ \phi_1^2\right) 
\end{pmatrix},
\end{align}
and similarly for $K_{,\phi_2}$.

The poles in the field-space metric in the new elliptic coordinates is more easily obtained
computing eq.\eq{KEij} in the new coordinates.  
The Einstein frame potential is 
$V_{\rm E} = V \cosh^2\rho/  {\sin^2 2\theta}$.
Pole inflation is obtained if $V$ is 
approximatively quadratic around the pole at $\theta=0$ and $\rho=\rho_0$
for (at least one) value of $\rho=\rho_0$.
In formul\ae, we need
\beq 
\label{eq:potentialmatrix}
V = R(\rho,\theta) \left[ c_{\theta\theta} \frac{\theta^2}{2} + c_{\rho\rho} \frac{(\rho-\rho_0)^2}{2}+
c_{\theta \rho} \theta (\rho-\rho_0)\right]\eeq
where $R$ is any smooth function around the pole. The $c_{ij}$ matrix determines the 
angle of incidence $\theta$ of the inflationary trajectory. 
Along it one gets a pole with order $p=2$ and residue
$\tilde\alpha_2 =3/2$,
and thereby the inflationary predictions of eq.\eq{nsrpred}.

 \smallskip

\subsection{Two-field pole inflation: elliptic singularity}
A similar model is obtained assuming $\xi_{1,2}<0$. 
The singularity curves are now ellipses, conveniently parametrised using  stretched polar coordinates $(\rho,\theta$)
\begin{align}
\label{ell1a}
\phi_1 =\frac{  \bp }{ \sqrt{-\xi_1}} \rho \sin \theta   ,
\qquad
\phi_2 =\frac{  \bp}{ \sqrt{-\xi_2}} \rho \cos \theta ,
\end{align}
so that $f=\bp^2(1-\rho^2)$.
The singularity curve  is $\rho = 1$, and we must expand around it in the $f\ge 0$ region.
The potential is $V_{\rm E} = V/(1-\rho^2)^2$.
In polar coordinates, the only element of the kinetic matrix with a pole around $\rho=1$ is
\beq K_{\rho\rho}^{\rm E}|_{\rm pole}=\frac{3\bp^2}{2(1-\rho)^2}
.\eeq
The resultant theory has once again attractor predictions given in eq.\eq{nsrpred}, and reheating proceeds in a similar way to the previous subsection.

\subsection{Two-field pole inflation: linear singularity}
We next examine inflation that features no inherent scale. Such a model will be further motivated in section~\ref{dimension-less}. In this case, 
we assume  $f=\xi_1 \phi_1^2 + \xi_2\phi_2^2$ with $\xi_1 = \xi_+>0$ and $\xi_2 =- \xi_-<0$.
The field-space curvature diverges at $\xi_1 (1+6\xi_1) \phi_1^2 + \xi_2(1+6\xi_2)\phi_2^2=0$,
away from the singularity curves, that are the two crossed lines 
$\phi_1/\phi_2= \pm \, \sqrt{\xi_-/\xi_+}$.
The best way to parametrise this theory is therefore a skew coordinate system
\begin{align}
\phi_\pm = \phi_1 \sqrt{\xi_+} \pm \phi_2 \sqrt{\xi_-}
\end{align}
so that $f=\phi_+ \phi_-$ and the two singularity curves are $\phi_+ = 0$ and $\phi_-=0$.
Due to the symmetry of the situation, we can focus on $\phi_-=0$.
This once again gives 
\begin{align}
K^{\rm E}_{++}|_{\rm pole} = \frac{3\bp^2}{2\phi_-^2}.
\end{align}
The resultant theory once again gives the attractor predictions of eq.\eq{nsrpred}.
As such, it is enough for the potential to vanish on at least one point on the singularity curve.
A potential with this property and which is exactly scale-invariant (quartic in the fields)
vanishes along all the singularity curve
 \begin{align}
 V = \frac{\lambda}{4}   \left( \xi_1 \phi_1 ^2 + \xi_2 \phi_2^2 \right)^2.
\end{align}

%

\subsection{Point singularity}\label{point}
We finally examine the situation where $f$ is not linear around the pole and/or
the singularity curve along which
$f=0$ has less than $N-1$ dimensions.

We consider a scale-invariant  theory with one field $\phi$ (see also~\cite{Cooper:1981byv}), so that 
\beq \label{eq:classical}
V(\phi) = \lambda \frac{\phi^4}{4},\qquad f(\phi)=\xi \phi^2\eeq
where $\xi$ and $\lambda$ are dimension-less coupling constants.
This leads to a 2nd order pole at $\phi=0$, $K_{\rm E} = \bp^2 (1+6\xi)/\xi\phi^2$.
The residue $\alpha_2/\bar M_{\rm Pl}^2= 6 + 1/\xi$
has a non-standard value because $f$ is quadratic (rather than linear)
around the pole at $\phi=0$.
The Einstein-frame potential is finite and exactly flat, $V_{\rm E} = \bar M_{\rm Pl}^4\lambda/4\xi^2$,
corresponding to $q=0$ in the language of section~\ref{overview}.
Planck-suppressed non-renormalisable operators in the potential could nonetheless still create a tilt and the desired SM minimum.

\smallskip

Let us next consider $N=2$ fields and a scale-invariant
$f=\xi_1 \phi_1^2 + \xi_2\phi_2^2$ with $\xi_{1,2}>0$.
Then $f=0$ only at the point at the origin in field space,  $\phi_{1}=\phi_2=0$.
It is convenient to use stretched polar coordinates 
\begin{align}
\label{ell1a}
\phi_1 =\frac{  \rho }{ \sqrt{\xi_1}}  \cos \theta   ,
\qquad
\phi_2 =\frac{  \rho}{ \sqrt{\xi_2}}  \sin \theta ,
\end{align}
such that $f=\rho^2$ is not linear around the pole at $\rho= 0$.
The dominant pole for the inflaton field  $\rho$ is again 2nd-order,
and its residual has the same non-standard value
\beq K_{\rho\rho}^{\rm E} = \frac{\bp^2 (1+6\xi)}{\rho^2\xi}\qquad\hbox{with}\qquad
\frac{1}{\xi}\equiv  \frac{\cos^2\theta}{\xi_1}+\frac{\sin^2\theta}{\xi_2}\eeq
obtained in the effective one-field case along a trajectory with angle $\theta$.
With more than $N=2$ scalars, pole inflation similarly arises from the origin of field space. 


\section{Pole inflation in dimension-less quantum theories}\label{dimension-less}
As discussed in the previous section, making gravity canonical can lead to
poles in the scalar kinetic terms, 
but pole inflation only arises if the potential $V$ satisfies non-generic conditions.
We already mentioned one possible rationale for them:
as the factor $f$ corresponds to a scale transformation,
nearly scale-invariant theories can have special properties.
We restrict the generic theory written in the Jordan frame in eq.\eq{Jordangeneric}
to field theories with dimension-less couplings only.


\smallskip

At quantum level, quantum corrections break classical scale invariance.
The quantum counterpart of eq.\eq{classical} 
is well approximated by the classical expressions with running couplings renormalised around the field value $\phi$
\beq V(\phi) =  \lambda(\phi) \frac{\phi^4}4,\qquad f (\phi)= \xi(\phi) \phi^2.\eeq
$K_{\rm E}(\phi)$ is no longer a simple power, and the Einstein-frame potential 
\beq
V_{\rm E}(\phi) = \frac{\bar M_{\rm Pl}^4}{4} \frac{\lambda(\phi)}{\xi^2(\phi)}\eeq
is no longer flat.
Roughly speaking, one can now have pole inflation with 
$p\approx 2$ and $q\approx 0$, where the small deviations from the classical
values arise from anomalous dimension.
More precisely, $V_{\rm E}(\phi)$ cannot be approximated as a simple function, 
unlike what is assumed in general treatments of pole inflation. 
The reason is that RG dynamics depend on $\ln \phi \propto \phi_{\rm E}$, the canonical scalar in eq.\eq{canoninfl}.

\smallskip

Since details of these functions matter, in this section we consider models that
describe the SM minimum too.
The scalar today sits at the SM minimum, $\phi=\phi_{0}$,
where $f(\phi_{0})=\bar{M}_{\rm Pl}^2$ and $V_{\rm E}(\phi_{0})=0$,
in view of the nearly vanishing of the cosmological constant.
The potential $V_{\rm E}$ features the desired SM minimum if the RG running is such that
\beq
\lambda(\phi_0)=0,\qquad  \beta_\lambda(\phi_0)=0,\qquad
\beta_{\beta_\lambda}(\phi_0)>0\eeq
at a field value $\phi_0$ such that $f(\phi_0)=\bar M_{\rm Pl}^2$~\cite{1403.4226}.
In the above equation $\beta_x \equiv d x/d\ln\bar\mu$ denotes the $\beta$ function of a generic coupling $x$,
and $\beta_{\beta_\lambda}$ denotes the $\beta$ function of the combination of couplings that arise in $\beta_\lambda$, the $\beta$ function of $\lambda$.
A similar notation is used below.
Since $\beta_{\beta_\lambda}\neq 0$, the tuning of parameters implied by $\beta_\lambda$ cannot be explained as a fixed point.

The slow-roll parameters of eq.\eq{srp} that determine inflation predictions
can be written in terms of the beta functions of the theory as
\begin{align}\label{sys:slowroll}
 \epsilon  &=   
 \frac{\xi}{1+6\xi}\frac12 \bigg[ \frac{\beta_{\lambda}}{\lambda} - 2 \frac{\beta_{\xi}}{\xi}\bigg]^2  ,\\
\eta &= 
\frac{\xi}{1+6\xi}\bigg[2\frac{\beta_{\beta_{\lambda}})}{\lambda} - 4\frac{\beta_{\beta_{\xi}}}{\xi}
-2 \frac{\beta^2_{\lambda}}{\lambda^2} +4\frac{\beta^2_{\xi}}{\xi^2}+
\frac{\beta_\xi/\xi}{1+6\xi}\left(\frac{\beta_{\lambda}}{\lambda} - 2 \frac{\beta_{\xi}}{\xi}\right)\bigg].
\end{align}
This shows that slow-roll inflation is generically achieved in  theories with perturbative dimension-less couplings.
More specifically $\epsilon,\eta \sim \lambda^2/(4\pi)^4$ contain a two loop suppression away from the potential minimum
where inflation ends and $\epsilon,\eta $ become large in view of the vanishing $\lambda$ at the denominator.


Since $p=2$, beyond the pole at the origin in field space $\phi=0$, 
another pole exists at $\phi=\infty$,
so that both small-field and large-field inflation can give acceptable predictions.

\subsection{Inflation with dimension-less theories at small coupling}
If couplings are so small that away from the minimum we have $|\epsilon|,|\eta|\ll 0.01$ (the size suggested by data about $n_s = 1-2\epsilon+\eta$), only the field region around the potential minimum matters. 
Then the potential can be expanded in Taylor series: at leading order the inflationary potential
is quadratic as function of the canonically normalised Einstein-frame scalar 
$\phi_{\rm E}(\phi)$, defined by $ {d\phi_{\rm E}}/{d\phi} = \sqrt{K}$. 
This implies that $N \lessapprox 60$ $e$-folds of inflation probe about 3 orders of magnitude of RG running
\beq\label{eq:runninginfl}
 \left| \ln \frac{\phi}{\phi_0} \right|\approx \sqrt{\frac{4\xi N}{1+6\xi}}   \lessapprox 6\eeq
and that the tensor/scalar ratio is $r \simeq 8/N \approx 0.15$~\cite{1403.4226}, above current bounds.\footnote{During inflation the Planck mass varies 
as $\bp = \sqrt{\xi}\phi \propto \exp[\pm \sqrt{2N/3}]$ for $\xi \gg 1$.
Large-field inflation starts from $\phi\ll \phi_0$ when the Planck length is much smaller,
giving a partial self-censorship of trans-Planckian inflation modes 
(conjectured to be a problem in~\cite{1909.11063}),
altought it risks involving trans-Planckian field values.
The opposite happens for small-field inflation.}
Including the next cubic order in the Taylor series can increase or reduce $r$~\cite{1502.01334}.
A substantial reduction down to acceptable values is possible; however
whenever cubics have an order unity effect 
(allowing to reduce $r$ down to currently acceptable values $r \sim 0.08$)
one expects that higher-order effects too start becoming relevant:
one can no longer use the Taylor expansion and needs a full theory.\footnote{If the full theory is
4-derivative gravity, it contains an extra inflaton candidate: the scalar present in the graviton in the presence of $R^2$ terms.
This gives a much smaller $r$, and dominates inflation if the Planckion $s$ is heavier than the scalar graviton.
Furthermore, a ghost lighter than the inflationary Hubble scale cancels the fluctuations in the graviton,
leading to a suppressed $r$~\cite{1703.08012}.
We here do not consider this possibility, and focus on pure QFT effects below the quantum gravity scale. }



\smallskip

In a full theory, $r$ gets significantly reduced (below the small-coupling limit that gives quadratic inflation)
if the couplings are such that the two-loop factor in eq.~(\ref{sys:slowroll})
gives the desired $|\epsilon|,|\eta| \sim 0.01$.
This corresponds to mild dimension-less couplings of order one.
More precisely, couplings must run by order unity in the 
$\approx 3$ orders of magnitudes probed by small-coupling
inflation according to eq.\eq{runninginfl}.

Sizeable couplings and a fast RG running come with the danger that couplings blow up,
running to strong coupling in IR (endangering computability and inflation) and/or in the UV (Landau poles presumably
ruin the consistency of the theory), unless 
an interacting fixed point in the UV (asymptotic safety) or in the IR is dynamically reached.

In theories where a scalar $\phi$ charged under some gauge interaction
has one quartic $\lambda$
and one Yukawa coupling $y$ 
(its presence allows to get a quantum potential with the desired SM minimum at $\phi_0$),
the one-loop RGE have the generic form\footnote{We do not here consider the gravitational sector,
where the dimension-less assumption leads to `agravity': a renormalisable 4-derivative action
with a possibly problematic ghost mode.}
\beq\begin{array}{rclrcl}\displaystyle
\frac{dg^2 }{dt}  &=& -b g^4, & \displaystyle
\frac{d\lambda}{dt} &=& \lambda (s_\lambda  \lambda + s_{\lambda y}  y^2 - s_{\lambda g}  g^2) -s_y y^4 + s_g  g^4 ,\\[2ex]
\displaystyle
\frac{d  y^2}{dt} &=& y^2(f_y  y^2 - f_g  g^2) ,& \displaystyle
\frac{d\xi}{dt} &=&\displaystyle \frac{1}{12}(1+6\xi)(\tilde{s}_\lambda  \lambda + s_{\lambda y}  y^2 - s_{\lambda g}  g^2),
\end{array}\eeq
with positive and gauge-independent order one coefficients $f_i$ and $s_i$, that depend on the specific model.
Notice the multiple appearance of $s_{\lambda y}$ and $s_{\lambda g}$.
We defined $t =\ln(E^2/E_0^2)/(4\pi)^2$ and 
included QFT effects and ignored quantum gravity effects (computable in agravity).

\smallskip

Let us count the number of free parameters:
\begin{itemize}
\item The SM minimum is generated if $\lambda(\phi_0)=\beta_\lambda(\phi_0)$ and $\beta_{\beta_\lambda(\phi_0)}>0$:
this fixes $\lambda$ and $y$ as $\lambda_0 =0$, $y_0^2/g_0^2 = \sqrt{s_g/s_\lambda}$ and restricts
$f_g  - f_y \sqrt{s_g/s_y} >b$
(the left side of the inequality is proportional to the sign of $\beta_y$ at $\phi_0$).
\item The condition $\xi_0 \phi_0^2 = \bar{M}_{\rm Pl}^2$ together with
the measured value of the amplitude of scalar perturbations
(see eq.\eq{infpredth} and\eq{infpredex})
can be used to fix $\xi$ and $\phi_0$:
assuming order one couplings one needs $\xi \sim 100$ and thereby sub-Planckian $\phi_0 =\bar{M}_{\rm Pl}/\sqrt{\xi}$.
Then the RG for $\xi_0 \gg 1$ implies it undergoes a nearly-multiplicative renormalisation,
with beta-function $\beta_\xi = (1+6\xi_0)(-s_{\lambda g} +s_{\lambda y}\sqrt{s_g/s_y})/12$ at $\phi_0$ 
that will be negative in all computed models.

\end{itemize}
In any given model, the gauge coupling $g$ remains as the only free parameter
(with small values giving quadratic inflation).
However, many models are possible and give different RG coefficients.


We next discuss the possible irreversible RG flows (limit cycles are considered impossible~\cite{1107.3987,1204.5221}):
to strong coupling in the UV (section~\ref{Landau}),
to strong coupling in the IR (section~\ref{confinement}), 
to an IR fixed-point (section~\ref{IRFP}), to an UV fixed-point (asymptotic safety, section~\ref{UVFP}).


\begin{figure}[t]
$$\includegraphics[width=0.42\textwidth]{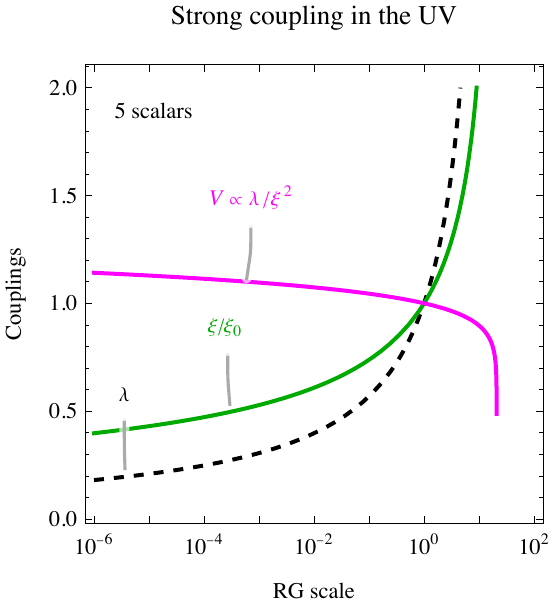}\qquad
\includegraphics[width=0.45\textwidth]{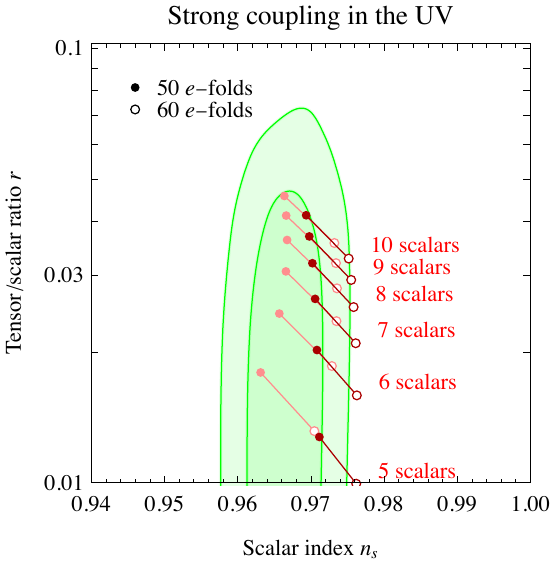}$$
\caption{\label{fig:sampleUV}\em Inflation with a scalar quartic
that becomes non-perturbative in the UV hitting a Landau pole.
{\bfseries Left}: sample of RG running in the model of section~\ref{Landau}.
{\bfseries Right}: its possible inflationary predictions, assuming small-field inflation that ends
at non-perturbative $\lambda=(4\pi)^2$ (dark red) or $10^3$ (light red).}
\end{figure}

\subsection{Inflation with strong coupling in the UV}\label{Landau}
If one wishes to consider RG running that leads to strong coupling at high energy
(and thereby to a Landau pole, possibly sub-Planckian)
the simplest theory has a Higgs-like scalar: 
$N$ ultra-light scalar real degrees of freedom $\phi_i$
with maximal global symmetry $\SO(N)$, so that
\beq f =  \phi_1^2+\cdots + \phi_N^2, \qquad V =\frac{\lambda}{4} ( \phi_1^2+\cdots + \phi_N^2)^2.\eeq
For example, the SM Higgs has $N=4$: the global symmetry arises accidentally due to
the $\SU(2)_L$ gauge symmetry 
and the Higgs is much lighter than the Planck scale for unclear reasons.
Assuming that gauge interactions are absent or negligible,
the one-loop RGE of the theory are
\beq (4\pi)^2\beta_\lambda = 2(8+N)\lambda^2, \qquad
(4\pi)^2\beta_\xi = (1+6\xi)\frac{2+N}{3}\lambda.\eeq
So $\lambda$ runs to small values in the IR, and $\xi+1/6$ is multiplicatively renormalised
\beq \lambda=\frac{\lambda_0}{1-2(8+N)t\lambda_0},\qquad
\xi+\frac{1}{6}=(\xi_0+\frac{1}{6})\left(\frac{\lambda}{\lambda_0}\right)^{(2+N)/(8+N)}.
\eeq
For $\xi \gg 1$, the Einstein-frame potential satisfies
$V_{\rm E} = V_{\rm E}^0 (\lambda/\lambda_0)^{(4-N)/(8+N)}$,
allowing small-field inflation for $N\ge 4$ and large-field inflation for $N<4$.
In the special case $N=4$, the quantum potential is flat up to terms suppressed by $\xi$.
Fig.\fig{sampleUV}a shows the running in the quasi-flat case $N=5$.

In this minimal model, inflation ends near to the Landau pole at strong coupling, 
where $\epsilon, \eta$ can become of order unity but the perturbative computation becomes unreliable.
Up to this caveat, fig.\fig{sampleUV}b shows the resulting predictions.
Like in pole inflation, we assumed small-field inflation out of the pole, and so
the predictions only hold assuming that 
the extra physics needed to generate the SM minimum starts to
be relevant after inflation end.
Otherwise, the predictions would be modified depending on the extra physics near to the SM minimum.
For example, one could add non-renormalisable operators, 
or an extra scalar, or extra couplings that appropriately modify the running.
In the next sections we will pursue the latter strategy,
considering less minimal models that allow for RG runnings
without Landau poles.


\begin{figure}[t]
$$\includegraphics[width=0.42\textwidth]{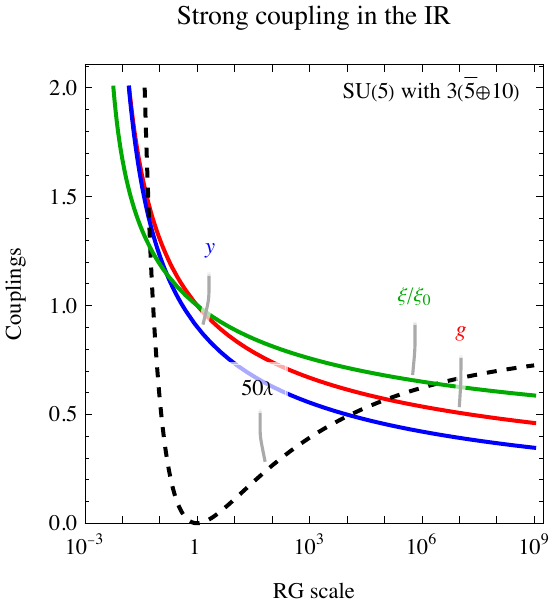}\qquad
\includegraphics[width=0.45\textwidth]{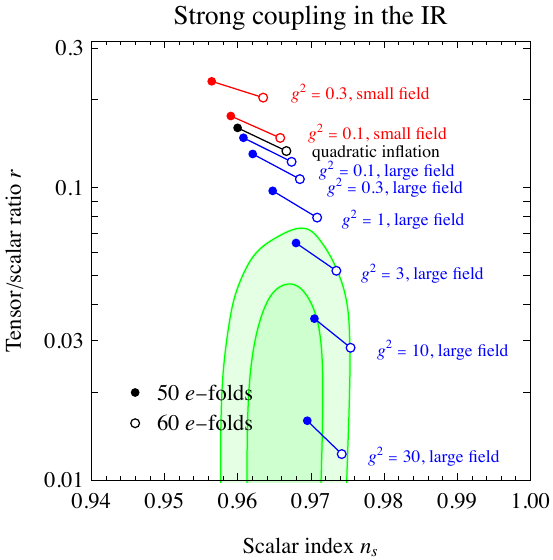}
$$
\caption{\label{fig:confinement}\em Inflation with couplings
that become non-perturbative in the IR giving rise to confinement.
{\bfseries Left}: sample of RG running in the model of section~\ref{confinement}
with $N_f=3$ and gauge coupling $g = 1$ at the SM minimum after inflation.
{\bfseries Right}: its inflationary predictions for different values of $g$.}
\end{figure}

\subsection{Inflation with strong coupling in the IR}\label{confinement}
In non-abelian gauge theories with a small matter content
(loosely speaking, fewer scalars and fermions than vectors),
the asymptotically free gauge coupling $g$ runs to non-perturbative values at some low-energy scale $\Lambda$,
analogously to QCD.
We add a charged scalar $\phi$ with a quartic coupling $\lambda$ that runs in a similar way.
Ignoring the issue of generating the SM minimum, large-field inflation would end at $\phi \sim \Lambda$.
We generate the SM minimum at a scale $\phi=\phi_0$
by adding fermions, such that one Yukawa coupling $y$ induces a running $\lambda$ with the desired 
SM minimum at an arbitrary scale $\phi_0$.
If $\phi_0$ is many orders of magnitude above $\Lambda$, 
the couplings $g(\phi_0)$ are weak and one gets the predictions of quadratic inflation. 
If $\phi_0$ is not much larger than $\Lambda$ so that couplings $g(\phi_0)$ are moderately large,
one gets different predictions.

\smallskip

As a concrete example, we consider a theory similar to the $\SU(5)$ extension of the SM:
a $\SU(5)$ gauge theory with $N_f$ generations of chiral Weyl fermions in the  anomaly-free
$\bar 5 \oplus 10$ representation
and a
complex scalar $\Phi$ in the fundamental 5 with quartic $\lambda |\Phi|^4$ and Yukawa coupling $y\,\Phi\,10\,10$.
With this coupling, order one factors in the RG equations 
\beq \begin{array}{ll}\displaystyle
(4\pi)^2\beta_g = -(\frac{109}{6} +\frac23 N_f)g^3,\qquad & \displaystyle
 (4\pi)^2\beta_\lambda = 12\lambda( 3\lambda+y^2 - \frac{12}{5} g^2) -6y^4+\frac{99}{25}g^4,\\[1ex]
 \displaystyle
 (4\pi)^2\beta_y = y(6 y^2-\frac{108}{5}g^2),& \displaystyle
 (4\pi)^2\beta_\xi = (1+6\xi)(4\lambda + y^2 - \frac{12}{5}g^2).
\end{array}\eeq
allow, if $N_f>1$, to generate the desired minimum 
in the Coleman-Weinberg $\Phi$ potential with a small enough Yukawa coupling 
(unfavourable order one factors would lead to a Landau pole in $y$ and/or to no minimum).
Fig.\fig{confinement}a shows a sample of RG running, 
with $y$ and $\lambda$ tuned in such a way that the Einstein potential has the desired SM minimum. 

As expected, for small couplings both small-field and large-field inflation reproduce the predictions of quadratic inflation,
with its too large $r\approx 0.15$.

Small-field inflation (from low energies $\phi< \phi_0$, in red in fig.\fig{confinement}b) 
leads to an excluded tensor/scalar ratio $r$,  because larger than the quadratic limit.
This is due to the larger value of the gauge coupling at low energies, and it is 
partially counter-acted by the fact that $\xi$ runs to larger values at lower energies.

The opposite happens for large-field inflation (from high energies $\phi>\phi_0$, in blue in fig.\fig{confinement}b).
Since couplings run to smaller values
at higher energy, a substantial reduction in $r$ needs a $\phi_0$ not much above $\Lambda$.
Different models would give qualitatively similar predictions (see for example fig.~3 of~\cite{1502.01334}).

\subsection{Inflation with an IR fixed-point}\label{IRFP}
Non-Abelian gauge theories with a large enough matter content
exhibit a qualitatively different behaviour:
the gauge coupling $g$ runs in the infra-red to a fixed-point $g=g_*$.
Banks and Zaks~\cite{Banks:1981nn} showed that
this phenomenon can be computed perturbatively in theories with
matter content such that the one-loop RG coefficient $-b$ is negative and accidentally small,
and its two-loop counterpart $b'$ is positive.
Then the 2 loop term in the RGE for $g$ is relevant,
\beq  \frac{d g^2}{d t}= - b g^4 + \frac{b'}{(4\pi)^2} g^6+\cdots,\qquad 
t=\frac{\ln \mu^2/\mu_*^2}{(4\pi)^2} \eeq
while higher order terms can be neglected, so that the IR fixed-point $g^2_* = (4\pi)^2b/b'$ is reliably computed.
The analytic solution to RGE where the fixed-point is approached for $\mu\lessapprox\mu_*$ is
\beq\label{eq:ggFPsol}
b t = \frac{1}{g^2}+\frac{1}{g_*^2} \ln \left(\frac{g_*^2}{g^2}-1\right)\qquad\hbox{i.e.}\qquad
g^2 = \frac{g_*^2}{1+W(e^{-1+b g_*^2 t})} 
\eeq
having introduced the Lambert function $W(z) e^{W(z)} = z$ to invert the relation between $g$ and $t$.
When $g$ reaches its IR fixed-point, the Yukawa $y$ and quartic coupling $\lambda$ behave as follows:
\begin{itemize}
\item $y$ too reaches an IR fixed-point, $y_*^2/g_*^2 = f_g/f_y$. 
\item If the condition $\beta_\lambda=0$ is satisfied at two values real of $\lambda$
(this depends on the values of the RG coefficients),
the higher solution is an IR-attractive fixed-point for $\lambda$.
If it is positive, the potential is acceptable.
In other words, dynamics self-adjusts the dimension of the $\phi^4$ composite operator to its classical dimension.
The fact that $\lambda(\bar \mu)$ remains finite at low energy means that our pole-inflation assumption $V(0)=0$ is satisfied.

\begin{figure}[t]
$$\includegraphics[width=0.42\textwidth]{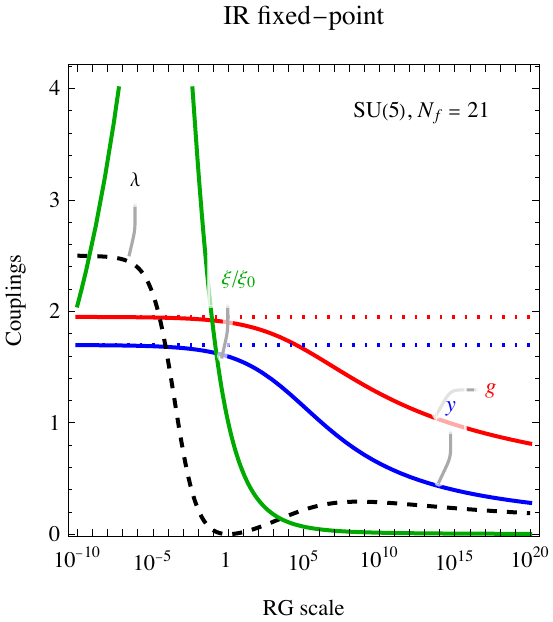}\qquad
\includegraphics[width=0.45\textwidth]{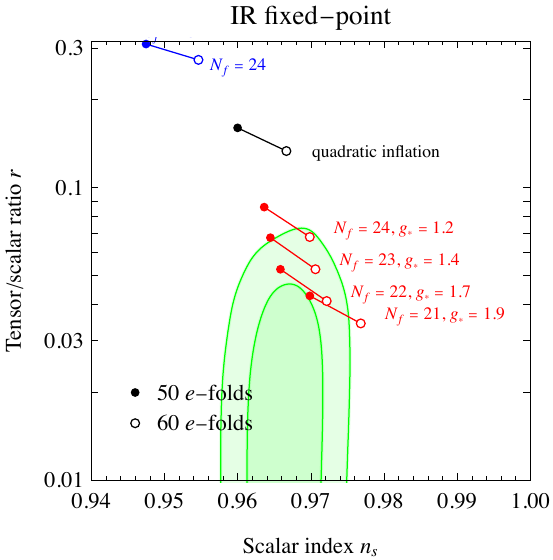}$$
\caption{\label{fig:sampleFPIR}\em Inflation with couplings
that approach a fixed point in the IR.
{\bfseries Left}: sample of RG running in the model of section~\ref{IRFP}
for $N_f=21$ corresponding to IR fixed-point $g_*\approx 1.9$.
{\bfseries Right}: its  small-field (red) and large-field (blue)
inflationary predictions for different values of $N_f$.}
\end{figure}

\item At the fixed-point for $\lambda, y$, the $\xi$ coupling gets renormalised multiplicatively.
In other words, the $\phi^2$ composite operator acquires an anomalous dimension.
\end{itemize}
We consider, as a simple example, a $\SU(N)$ gauge theory with $N_f$ Weyl 
fermions $\psi_i \oplus \psi^c_i$ in the $N\oplus\bar N$,
one fermion singlet $\nu$, one complex scalar $\Phi$ in the fundamental $N$.
Without loss of generality, the Yukawa interactions and the potential can be written as
\beq \label{eq:YukV}
\Lag_{\rm Yuk}=y\, \Phi \psi_1^c \nu + \bar y \,\Phi^* \psi_1\nu+\hbox{h.c.}, \qquad
V  = \lambda  |\Phi|^4.\eeq
The RG equations are

\begin{align}\label{sys:RGEmodel}
  \frac{dg^2 }{dt}  &=   -b g^4 + \frac{1}{(4\pi)^2}
\left[\frac{ \left(\left(13 N^2-3\right) N_f-34 N^3+4 N^2-3\right)}{3 N}g^6 -\frac{1}{2} g^4 \left(\bar{y}^2+y^2\right)\right] 
 , \\
 \frac{d  y^2}{dt} &=   y^2\left[\frac{3+N}{2}  y^2 +\frac{6+N}{2}- \frac{3(N^2-1)}{2N}  g^2\right] ,  \hspace{10em}
 \\
  \frac{d\lambda}{dt} &=  \lambda  \left[2   \lambda  (N+4) +
2 (y^2+\bar{y}^2)+3\tfrac{N^2-1}{N}g^2 \right]-( y^2+ \bar{y}^2)^2
+3 g^4 \tfrac{N \left(N^2+N-4\right)+2}{8 N^2},\\
  \frac{d\xi}{dt} &=  (6 \xi +1) \left[\frac{1}{6} \left(\bar{y}^2+y^2\right)-\frac{N^2-1}{4 N}g^2 +\frac{N+1}{3} \lambda\right],
\end{align}
where $b=\frac{11}{3}N-\frac23 N_f - \frac16 $ is chosen small and positive, and a similar RG holds for $\bar y$.
The perturbative IR fixed point for $g$ exists for $N_f \approx 11N/2$;
it is lost for higher $N_f$ and perturbative control is lost for lower $N_f$
(we are interested in mildly large couplings).
Given the order one factors, the acceptable parameter space of the model is somehow restricted.
Assuming, for simplicity, $\bar y = y$, the value of the Yukawa coupling
$y$ needed to have the desired SM minimum in the SM minimum
obtained imposing $\lambda=\beta_\lambda=0$ is below the fixed-point value of $y$
for $3<N<12$ by less than $4\%$ (larger values of $y$ lead to a Landau pole).
Furthermore, the gauge coupling $g$ at the SM minimum
where $\lambda=\beta_\lambda=0$ must be similarly near to its fixed point
to avoid having maximum rather than a minimum,
as can be seen by imposing $\beta_{\beta_\lambda}>0$.

Fig.\fig{sampleFPIR}a shows a sample of RG running assuming $N=5$ colours and $N_f=21$ flavours. 
We choose a somehow low $N_f$ such that the Banks-Zaks fixed-point has a moderately large coupling $g_*$,
in order to obtain inflationary predictions that substantially deviate from 
those of quadratic inflation obtained in the weak coupling limit.
Fig.\fig{sampleFPIR}b shows that small-field inflation 
(from the pole at $\phi=0$) can give acceptable predictions for $n_s$ and $r$.
On the other hand large-field inflation predicts a too large $r$, if
QFT RG evolution near to the Planck mass can be trusted.

%
%




\begin{figure}[t]
$$\includegraphics[width=0.42\textwidth]{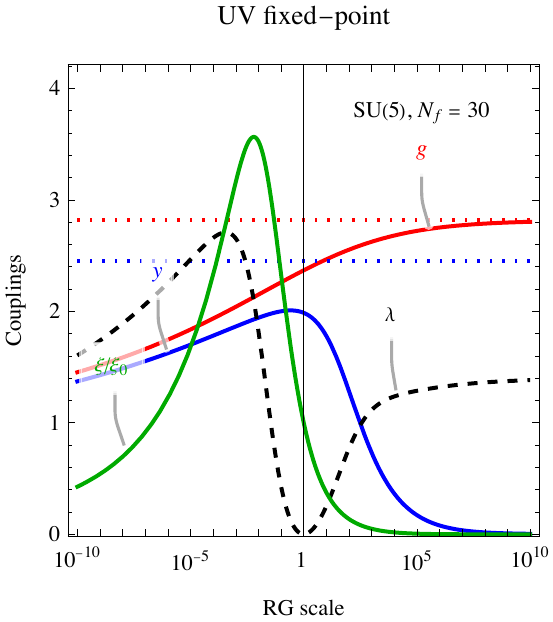}
\qquad
\includegraphics[width=0.45\textwidth]{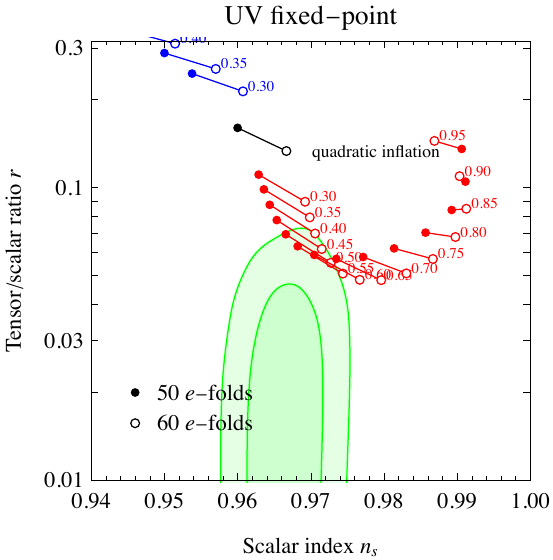}$$
\caption{\label{fig:sampleFPUV}\em Left: Inflation with couplings
that approach an interacting fixed point in the UV.
{\bfseries Left}: sample of RG running in the model of section~\ref{UVFP} with $N=5$ and $N_f=29$.
{\bfseries Right}: its  small-field (red) and large-field (blue)
inflationary predictions for different values of $g_0/g_*$.}
\end{figure}

\subsection{Inflation with an UV fixed-point}\label{UVFP}
As we are interested in moderately large couplings that avoid Landau poles,
the remaining possibility is asymptotic safety: dimension-less couplings that run up to an interacting UV fixed-point.
Litim and Sannino~\cite{1406.2337} found that this is perturbatively realised in specific models,
where the one-loop gauge beta function coefficient $b$ is small and negative,
and the two-loop term is positive thanks to the contribution of a Yukawa coupling $y'$ that reaches its UV fixed-point.
As a result the gauge coupling runs according to eq.\eq{ggFPsol}, where now $b<0$.
The models involve a $\SU(N)$ gauge theory with $N_f$ Weyl fermions in $N \oplus\bar N$ 
and $N_F^2$ neutral scalars $S_{ij}$ with (for simplicity) a common Yukawa $y\,S_{ij} \psi_i \psi_j^c$.
Since the scalars $S_{ij}$ are neutral, their scalar potential does not give rise to the desired 
post-inflationary SM minimum.

Following~\cite{1701.01453}, we thereby add an extra inflation-candidate
scalar $\Phi$ in the fundamental of $\SU(N)$, with a Yukawa coupling to one fermion flavour,
as in eq.\eq{YukV}.
We can assume, for simplicity, that the mixed scalar quartic  $|\Phi|^2 |S_{ij}|^2$ vanishes.
Then the RG equations for $y,\lambda, \xi$ are as in eq.s~(\ref{sys:RGEmodel}).

Fig.\fig{sampleFPUV}a shows an example of the running
tuned in such a way that the Einstein potential has
the desired SM minimum at $\phi=\phi_0$ where $g=g_0$.
We assumed $N=5$ colours, $N_f=30$ flavours
(such that the UV fixed-point of $g_*$ is moderately large, allowing for significant deviations
from the weak coupling limit), $\bar y = y$ and $g_0/g_*=0.5$.
Fig.\fig{sampleFPUV}b shows the resulting inflationary predictions as function of $g_0/g_*$.
Large-field inflation (in blue) gives too large $r$,
small-field inflation (in red) allows acceptable values of $r$ but not below $r=0.05$ because
when $g_0$ gets so large that details of the theory matter,
the predicted $n_s$ and $r$ start deviating from the best-fit  values.
A similar result is found for different number of colours $N$.


%

\section{Conclusions}\label{concl}
A pole in the kinetic terms of one scalar $\phi$ leads to pole inflation.
We explored the possibility that the same phenomenon arises in theories without poles in the Jordan frame, but rather in the Einstein frame originating
from a non-minimal coupling of a scalar $f(\phi)$ to gravity. 
Points in field space where $f=0$ (namely, where gravity is strongly coupled) lead to pole inflation
provided that the Jordan-frame scalar potential vanishes fast enough at the pole.
In this case, pole inflation is recovered through the usual ``stretching'' of the canonicalised field. 
The needed condition on the potential is non-generic, and similar to the vanishing of the cosmological constant at the SM minimum.

\smallskip

We next explored the generic case with $N$ scalars $\phi^i$, where we found that
the pole is generically promoted to a singularity curve, the ($N-1$)-dimensional surface where
the  non-minimal coupling vanishes, $f(\phi_i)=0$.
We describe the required form of the Jordan frame potential such that the multi-field generalisation of pole inflation is realised: the potential needs to vanish fast enough around at least one point of this surface. 
The field space is now generically warped, but the curvature is generically finite at and near the singularity surface
and has negligible effects.
The inflationary trajectory asymptotically approaches the singularity surface with an angle of incidence that depends on the potential.
This modifies the effective residue at the pole.
Up to this caveat, one obtains
the usual attractor predictions that depend solely on the order and the residue of the pole. 
The pole has 2nd order and a universal residue whenever $f(\phi_i)$ is approximatively linear around $f=0$.
Assuming an approximately linear tilt in the potential, this gives the same predictions as Starobinsky inflation.

At first glance, the above situation has two unsatisfactory aspects:
\begin{enumerate}
\item The predictions of pole inflation arise assuming that inflation is fully dominated by field
values around the pole, so that the structure of the potential around the SM minimum is neglected.

\item Canonicalisation of the graviton (namely, going from a generic Jordan frame to the Einstein frame) 
gives rise to pole inflation only if the potential has a non-generic structure.
\end{enumerate}
In section~\ref{dimension-less} we showed that the needed structure of the potential
automatically arises in quantum theories with dimension-less couplings only.
A small tilt is provided by their RG running that lifts the exact flatness of the classical potential.
Having a physical origin for the tilt, we can thereby abandon the assumption that 
inflation is fully dominated by the pole, and consider theories where the SM minimum is present and
affects inflationary predictions.
Small couplings universally lead to slow RG running and to nearly-quadratic inflation,
that predicts a too large tensor/scalar ratio.
We thereby consider larger couplings that provide
the possible faster RG running that can arise in QFT:
a coupling that runs either to a perturbative fixed point or to non-perturbative values,
either in the UV (figures\fig{sampleUV},\fig{sampleFPUV})
or in the IR (figures\fig{confinement},\fig{sampleFPIR}).
We then found that some of these models can provide acceptable inflationary predictions. 
In all the considered models, RG coefficients are such that the
non-minimal coupling $\xi$ to gravity runs to smaller values at higher scales.
Different models where $\xi$ runs in the opposite way could give different predictions.

\paragraph{Acknowledgements}
This work was supported by the ERC grant 669668 NEO-NAT.

\end{document}